# NEW HORIZONS: LONG-RANGE KUIPER BELT TARGETS OBSERVED BY THE HUBBLE SPACE TELESCOPE

S. D. Benecchi[1,2], K. S. Noll[3], H. A. Weaver[4], J. R. Spencer[5], S. A. Stern[5], M. W. Buie[5] and A. H. Parker[6]

[1] Planetary Science Institute, 1700 East Fort Lowell, Suite 106, Tucson, AZ 85719; susank@psi.edu
[2] Carnegie Institution of Washington, Department of Terrestrial Magnetism, 5241 Broad Branch Road, NW, Washington, DC 20015.
[3] NASA Goddard Space Fight Center, 8800 Greenbelt Rd. Code 693, Greenbelt, MD 20771
[4] Space Department, Johns Hopkins University Applied Physics Laboratory, 11100 Johns Hopkins Road, Laurel, MD 20723.
[5] Southwest Research Institute, 1050 Walnut St., Suite 300, Boulder, CO 80302.
[6] Department of Astronomy, University of California at Berkeley, B-20 Hearst Field Annex #3411, Berkeley, CA 94720.

**Proposed running head:** New Horizons Long-Range Observation KBO Targets

**Corresponding author:** S. D. Benecchi, susank@psi.edu, Planetary Science Institute, 1700 East Fort Lowell, Suite 106, Tucson, AZ 85719.

**ABSTRACT**

We report on Hubble Space Telescope (HST) observations of three Kuiper Belt Objects (KBOs), discovered in our dedicated ground-based search campaign, that are candidates for long-range observations from the New Horizons spacecraft: 2011 $JY_{31}$, 2011 $HZ_{102}$, and 2013 $LU_{35}$. Astrometry with HST enables both current and future critical accuracy improvements for orbit precision, required for possible New Horizons observations, beyond what can be obtained from the ground. Photometric colors of all three objects are red, typical of the Cold Classical dynamical population within which they reside; they are also the faintest KBOs to have had their colors measured. None are observed to be binary with HST above separations of ~0.02 arcsec (~700 km at 44 AU) and $\Delta m \leq 0.5$.

# 1. INTRODUCTION

The New Horizons spacecraft, launched in January 2006, will fly through the Pluto system in July 2015. The Pluto system, consisting of a large binary ($d_{Pluto}$ = 2306 km, and $d_{Charon}$ = 1307 km) with four smaller satellites ($d_{satellites}$~ 10 - 170 km), is one of the largest and most complex members of the Kuiper Belt (Jewitt & Luu, 1993). This region has more than 1700 objects that have been discovered and cataloged, and is estimated to contain ~$10^5$ objects having diameters larger than $d$~100 km (Petit et al. 2011). The physical properties of smaller ($d$~30-300 km), less atypical Kuiper Belt Objects (KBOs) provide important points of comparison and context for understanding the Pluto system and its relationship to the larger Kuiper Belt. A flyby of a KBO and studies of other encounterable objects have always been key goals for the New Horizons mission beyond the Pluto encounter as identified for the NASA KBO/Pluto mission (which became New Horizons) in the 2002 Planetary Decadal Survey (Belton et al. 2002). The New Horizons spacecraft and mission was designed, therefore, to allow redirection to one or more KBOs between 2017 and 2021 (Stern 2008; Spencer et al. 2003) provided that a suitable candidate can be identified. This is likely to be the *only* opportunity in our lifetimes to encounter a KBO other than Pluto.

Because precise orbits for flyby and long-range candidates are critical to the success of the New Horizons extended mission, we present in this paper Hubble Space Telescope (HST) follow-up observations of candidate targets exploiting HST's sensitivity, resolution, and PSF stability. These HST observations also allow us to make basic color measurements and to search for binary companions down to ~0.02 arcseconds in angular separation at the time of the observations.

## *1.1. Identifying Potentially Encounterable KBOs*

The KBOs that are potentially encounterable by New Horizons are limited to those with orbits that place them within reach of the remaining $\Delta$v capability of the spacecraft, currently estimated to be $\Delta v \approx 130$ m $s^{-1}$. This restriction means that by far the most likely candidate target(s) will be a member of the low-eccentricity, low-inclination Cold Classical subpopulation. In 2004 a ground-based search for bright objects ($m_R \geq 24.0$) was conducted over a wide area of sky (a few degrees) through which objects that will be in the New Horizons spacecraft trajectory during the optimal KBO encounter dates were located. This is defined by projecting the nominal trajectory of New Horizons after the Pluto encounter on the sky, then tracking the orbits of those potential KBOs backward (and forward) in time and orbital space. The total sky area that these potential KBO orbits cover is larger the farther out in time from encounter one searches. The closer the search time is to encounter, the more collapsed the potential KBO orbits are in the sky search area, however, this also means a shorter astrometry baseline for accurate obit determination and targeting of the discovered KBO. Unfortunately, no bright accessible object was found in the initial survey (Buie, 2012). In 2011, when ground-based searching conditions became optimized – the search area was small enough to justify the use of significant amounts of observing time on large telescopes – a deep search campaign (ideally to $m_R$ ~ 26.5, but more realistically to $m_R$ ~26.0,) was begun in earnest. As of January 2014, we have utilized over 520 hours (~52 nights) of telescope time on large aperture facilities (the 8.4-m Subaru telescope with SuprimeCam and the 6.5-m Magellan telescopes with the MegaCam and IMACS cameras; Buie et al. 2012; Spencer et al. 2013). No KBO has been found yet that is within the Δv fly-by capabilities of New Horizons.

Identifying encounterable KBOs has been especially challenging because the search region is near the galactic plane where the high star density and consequent confusion limit the depth to which we can effectively search. Additionally, there appears to be a turnover in the KBO magnitude distribution at small (faint) sizes (Fraser et al. 2014) that is more severe than originally anticipated. It is desirable to detect potential encounter objects as early as possible because the uncertainty in the predicted location of the KBO encounter candidate depends critically on the of the orbital arc length of the object. Some of the phase space where KBO encounter objects could be found is lost if retargeting of the spacecraft cannot be done shortly after the Pluto system encounter. A minimum of a 1-year arc on the KBO orbit is necessary for the initial burn, with continued orbit refinement closer to the time of the KBO fly-by maneuver itself.

However, a targeted close fly-by of an object is not the only New Horizons KBO science available. Longer range observations of objects, passing within 0.3 AU of New Horizons that are bright enough to image from the spacecraft with ~150 km resolution or ~5x HST resolution, provide opportunities for measurements with the New Horizons instruments that exceed the capabilities of Earth-based facilities and that provide unique phase angle coverage not available for any objects observed from the Earth. At the time of this paper our team has discovered 52 KBOs including a high inclination Neptune Trojan 2011 $HM_{102}$ (Parker et al. 2013a). At least five of these objects were long-range observation candidates (<0.3 AU) with ground-based orbital arcs sufficient for HST targeting, including the four objects presented in this paper (Table 1). Follow-up observations of these objects were prioritized by their magnitudes, orbital arcs and an evaluation of their encounter probabilities based on Monte Carlo models of their projected orbits with respect to the New Horizons spacecraft trajectory and the Δv available to the spacecraft for maneuvering. Four additional objects among our discoveries, but not presented here, were found to have preliminary predictions to pass within long range targeting by New Horizons, however, they were not recovered enough times in our ground-based data to extend their orbital arcs beyond a few nights to confirm their long range observation candidacy and to guarantee observation within the HST field of view. In most cases, their magnitudes were also significantly uncertain. Details of our ground-based campaign will be presented in a future paper.

INSERT TABLE 1 HERE

## 2. OBSERVATIONS

Observations were obtained using the Wide Field Camera 3 (WFC3) in the UVIS channel on HST (0.04 arcsec/pixel, field of view 162 arcsec square). Objects were tracked at the rate of motion of the KBO and observations were made using the F606W filter, nominally centered at 600.1 nm with FWHM of 150.2 nm, comparable to a Johnson V filter and the F814W filter, centered at 799.6 nm with FWHM 152.2 nm, comparable to Cousins I (Heyer et al. 2004). The V-I colors (or the equivalent colors in HST filters) of the candidate objects can determine if they have the red colors characteristic of Cold Classical KBOs (Gulbis et al. 2006; Peixinho et al. 2008). If we found an object with a neutral or blue color, it might potentially be an interloper from another dynamical class of the Kuiper Belt. With multiple candidate targets for distant observations with New Horizons, objects with a spread of colors might provide the most useful scientific insight to the properties of the Kuiper Belt region if in fact their surfaces sampled different populations in the Kuiper Belt. In Cycle 19, September 2012, we observed three of the best long-range candidates (HST 12535: 2011 $JW_{31}$, 2011 $JY_{31}$, and 2011 $HZ_{102}$) and in Cycle 20, September 2013, we observed the newly discovered long-range candidate 2013 $LU_{35}$ (HST 12887). For each object a pair of HST orbits were utilized, obtaining both astrometric and photometric measurements.

Observations for each orbit followed the sequence: F606W – F814W – F814W – F606W with small dithers between each exposure and with 1-15 HST orbit separation, so that the KBOs could be seen to move relative to the background star field between orbits. Exposures were 455 seconds in F606W and 640 seconds in F814W with the goal of obtaining S/N~5 for a single image, and color measurements good to ~0.1 magnitudes with the combined images. Table 2 summarizes the raw observations. Because these objects are in high density star fields, HST observation times were pre-selected based on the best known star catalogs to limit the probability of being close to, or overlapping, bright stars in the field of view. A sample HST image for each object in a single F606W integration are given in Figure 1.

INSERT FIGURE 1 HERE

INSERT TABLE 2 HERE

## 3. DETECTION, PHOTOMETRIC AND ASTROMETRIC ANALYSIS

### *3.1. Detection and Photometry*

To identify the KBOs of interest in our fields, the HST images were registered and then stacked based on the expected plane of sky motion of the KBO. Registration offsets are non-linear due to the parallax from HST's orbit around the Earth. This process led to the recovery of three of the four KBOs observed. One of the targeted KBOs, 2011 $JW_{31}$, was not found. The expected recovery area was unluckily in a cluster of fainter stars not bright enough to be visible in the field pre-selection images. It was also equally the faintest of our first three original sources (Cycle 19: $m_{2011\ JY31}$ ~ 24.7, $m_{2011\ JW31}$ ~ 25.4, $m_{2011\ HZ102}$ ~ 25.4; Cycle 20: $m_{2013\ LU35}$ ~ 26.0, with uncertainties as large as a few tenths of a magnitude). As pointed out in Benecchi et al. 2013, KBOs in this size range often exhibit substantial lightcurves, varying up to a full magnitude, on relatively short timescales (~8 hours or less). The combination of these factors may have led to the non-detection of this object, or perhaps the ephemeris was not as good as predicted.

The other three objects were detected in the stacked images. Once identified it was possible to identify them in individual images. In a few individual images, the KBOs were close to bright stars or impacted by a cosmic ray or bad pixel. Photometry was carried out using a small aperture on both the KBO and nearby close stars with clean backgrounds. Because KBOs move between frames relative to the background stars, for the few KBO frames with star contamination, the photometry of the star was measured on images when the KBO was not overlapping. On the frames with the overlapping sources, the combined source was measured and the star counts from the other frame were subtracted out to give just the counts for the KBO on the contaminated frame (with slightly larger uncertainties). In one case cosmic ray contamination in a single pixel was corrected by replacing that pixel with the average of the neighboring pixels, again with slightly larger uncertainties on the resulting magnitude. The background sky was determined using the Goddard Space Flight Center (GSFC) IDL *sky* routine, part of the IDL Astronomy Library (http://idlastro.gsfc.nasa.gov), in all cases.

The calibrated fluxes, expressed in the Vega magnitude system (VEGMAG), were calculated from the observed counts using the inverse sensitivity and photometric zeropoint keyword values (PHOTFLAM and PHOTZPT) from the HST image headers (Rajan et al. 2011). We used the synthetic IRAF photometry routine *synphot* to convert from the STMAG to the VEGAMAG system. F606W and F814W average magnitudes for each HST orbit were then found and the larger of either the combined uncertainties or the scatter in the values (2 data points per orbit per filter, 4 per object per filter for the pair of orbits) were used to set the magnitude uncertainties. Each visit also provided a color measurement that was then averaged between the two HST visits for an overall color determination with the same procedure as in the individual orbits used to set the uncertainties. The photometric and color results can be found in Tables 2 & 3.

INSERT TABLE 3 HERE

### *3.2. Astrometry*

Our observations confirm that ground-based discovery ephemerides with arcs of ~30 days are sufficient for recovery in the WFC3 field of view up to a year later. To provide a reference point, for an object such as 2011 $JY_{31}$, which has an orbital period of 292 years, the original arc is significantly improved at each interval where astrometry is measured and as the orbital arc-length grows the positional uncertainty shrinks rapidly as can be demonstrated by modeling (Wasserman et al. 2006) using orbit codes such as Bernstein and Khushalani (2000). The discovery arc, typically 24 hours, covers only 0.0009% of the orbit. We use the Minor Planet Center (MPC) astrometric record of 2011 $JY_{31}$ for illustration (some of the measurements included in the values for Table 1 have yet to be reported to the MPC). This object was discovered on 18 April 2011 and re-observed 6 days, 34 days, 65 days, 1 year and 1.4 years later. With a 6 day arc the total error on the object position is 0.2 arcseconds, or 736 arcseconds if propagated a year forward. At each progressive interval the immediate error falls to a tenth of an arcsecond, while the propagated errors are 42, 40, 11 and 0.3 arcseconds, respectively. For our HST observations which occurred ~510 days (1.4 years) from discovery (the final astrometric point previously mentioned) with a ground-based arc of 1 year, the object was recovered within ~10 HST pixels (~0.4 arcsec) of the predicted location. The other recovered objects had similar recovery and recovered properties.

The accuracy of the astrometry also plays a key role in targeting precision since uncertainties in the stellar positions translate to uncertainties in the true position of the KBO at a given time. Relative astrometry from HST for faint targets such as these have a precision of ~1 pixel or 0.04 arcseconds. Absolute astrometry is accurate to 200 mas, limited by the accuracy of astrometric catalogs used to establish the world coordinate system in HST images. For each HST image, astrometry was carried out following the method of Benecchi et al. (2010). The astrometric results, including the geocentric position of HST at the time of observation, were reported to the Minor Planet Center for improved ephemerides.

Future improved global astrometric catalogs either from ground-based surveys using facilities like the CFHT or in the long term with astrometry from the GAIA mission will allow ground-based and HST astrometry to be redone for improved network and targeting accuracy for fly-by or long range observations by New Horizons. Stellar astrometry from GAIA, expected to be at the level of 24 microacrseconds and anticipated to be available within a year or two of GAIA's launch, but well before final corrections would be made for a KBO long-range observation or encounter, will improve our results by ~2 orders of magnitude (Buie, private communication). Preliminary calculations suggest that a 3-year baseline orbit can be retroactively improved to the equivalent precision of a 34-year baseline (12% of a typical ~275 year orbit period for a KBO with *a*=42 AU), making KBO fly-by or long-range observation planning considerably easier.

## 4. BINARITY, COLOR AND SIZE

Our HST observations allow us the ability to 'pre-classify' potential long range and encounter candidates in order to prioritize the scientific return for the New Horizons spacecraft observations. The New Horizons spacecraft's trajectory through the Pluto system sends it through the region of space populated by the low-inclination, low-eccentricity Cold Classical Kuiper Belt. This is a population that studies show is distinctive both dynamically and photometrically from all other minor planet populations (Elliot et al. 2005; Fraser et al. 2012) and may be the most primordial population of objects that can be observed in the Solar System. Oort cloud comets, which reside farther out, are believed to have formed closer to the Sun and then ejected outward, so while they are in cold storage, they are likely to be more processed than the KBOs in the fly-by region. Cold Classical KBOs are thought to have formed near their current location and not to have been substantially scattered during the giant planet migrations (Parker & Kavelaars 2010; Batygin et al. 2011; Volk & Malhotra 2011). These KBOs are found to be statistically redder than other populations (Gulbis et al. 2006; Peixinho et al. 2008; Benecchi et al. 2011), to have high albedos (Brucker et al. 2009) and $29^{+7}_{-6}$% (Noll et al. 2008) were found to be binary (although we don't know if this binary rate holds for smaller objects, Nesvory et al. 2011).

To address the question of binary systems, we examined the individual KBO PSFs to look for evidence of binarity. In all cases, the results from HST were negative. No companions are detected with separations of ~0.02 arcsec (~700 km at 44 AU) and $\Delta m \leq 0.5$. Closer undetected companions may yet be detectable with long-range observations from New Horizons where the angular resolution exceeds that of HST (Parker et al. 2013b).

The observed colors of the three new KBOs are shown in Figure 2 where we compare the newly obtained data with 142 KBO colors that have been measured in the same HST filters (Benecchi et al. 2009; Fraser et al. 2012). The F606W-F814W colors (a proxy for V-I) plotted span a wide range from 0.5 < F606W-F814W < 1.3. The three newly detected KBOs are at the red end of the range with F606W-F814W>0.75 (including the moderately large 1σ uncertainties). The red colors are consistent with the expectation that most of the targets in the search area will be members of the Cold Classical population. All three long-range New Horizons KBO observation candidates observed by HST are significantly fainter than the other KBOs plotted here and are among the faintest to have had their colors measured.

INSERT FIGURE 2 HERE

Using the formalism of Bowell et al. (1989) and an assumed geometric albedo of ρ=0.1, our long-range targets have diameters that range from 50-100 km. These objects are in the approximate size range where sizable amplitude lightcurves (>0.2 magnitudes) might be expected based on results from Trilling et al. (2006) and Benecchi & Sheppard (2013). Absolute variations in magnitude for these objects during a single HST orbit and between pairs of orbits are consistent with the uncertainties in our measurements (~0.1 magnitudes) so at least at our sparsely sampled intervals we don't find these objects to have large amplitude lightcurves indicative of highly elongated shapes.

## 5. SCIENCE ENABLED BY LONG RANGE AND ENCOUNTER KBO OBSERVATIONS FROM NEW HORIZONS

Combining our HST results with ground-based detection limits, we expect that the New Horizons spacecraft encounter object will likely be fainter than our current ground-based delectability limit of $m_R$=26.0 in good seeing (0.45 arcsec) over a small fraction of the search area and limit of $m_E$=25.6 over 75% of the search area, giving it a likely diameter of ≤ 65 km (assuming an albedo of ρ=0.1; Bowell et al. 1989). A higher albedo, not unexpected for objects in this subpopulation of the Kuiper Belt (Brucker et al. 2009), corresponds to a smaller object. Schlichting et al. (2013), in comparative studies of the observed KBO size distribution with theoretical models, suggest that a diameter of ~60 km is an important size threshold for studying collisional processing in the Kuiper Belt. They propose that objects above this size are primordial (unmodified by collisional evolution over the age of the Solar System) while smaller objects have been modified by collisions. Therefore, observations of both the size distribution of KBOs in this size regime, as well as high resolution studies of these objects with the New Horizons spacecraft with either long range observations or a close fly-by (Table 1 provides estimated observational resolution for each long-range observation candidate), can provide valuable knowledge about the growth and destruction of large KBOs.

For a likely encounter distance (10000-20000 km), New Horizon's instruments can provide broadband and multicolor images with resolutions of ~100 and ~400 meters, respectively; near-IR spectral maps with ~2 km resolution sensitive to volatiles such as water ice, methane ice and ammonia; and UV spectra and charged-particle data sensitive to active volatile loss, in addition to concurrent heliocentric plasma and dust-particle measurements (Weaver et al. 2008). Comparison of the characteristics of the Pluto system, thought to be formed through a large collision (Canup 2011), with studies of the anticipated 'small' KBO to be encountered by New Horizons, will provide a sample of the two different size regimes of the Kuiper Belt population (Stern & Spencer, 2003), which have very different formation and evolutional histories.

## 6. CONCLUSIONS

We have reported on the basic photometric characteristics of three KBOs that are observable from the New Horizons spacecraft at long range (<0.3 AU) after the Pluto system fly-by science has been completed. We do not identify any of these objects as binary with HST at the level of 0.02 arcsec separation (~700 km) and Δm<0.5. However, the region in which New Horizons will fly is near the center of the Cold Classical Kuiper Belt where nearly 30% of objects are binary (Noll et al. 2008) and at <0.3 AU New Horizons has ~5x the resolution and sensitivity of HST and could detect companions unresolvable from HST. The three newly discovered objects are all red in color with 0.75 < F606W-F814W < 1.2 consistent with being members of the Cold Classical population. Distant flybys will enable photometry at phase angles unobservable from the Earth that can eventually be cross-calibrated with the observations at low phase angle reported here. The utility of HST astrometry from our measurements can be improved by future upgrades to absolute astrometry from improved ground (CFHT) or space-based (GAIA) stellar astrometry so that these objects can be targeted accurately with the New Horizons spacecraft for long range observation, with better resolution than any Earth based facility.

## ACKNOWLEDGMENTS

This paper is based in part on ground-based data collected at the Subaru Telescope, which is operated by the National Astronomical Observatory of Japan, and on data gathered with the 6.5 meter Magellan Telescopes located at Las Campanas Observatory, Chile. Space-based observations were made with the NASA/ESA Hubble Space Telescope, obtained at the Space Telescope Science Institute, which is operated by the Association of Universities for Research in Astronomy, Inc., under NASA contract NAS 5-26555. These observations are associated with programs 12535 and 12887. Support for both programs were provided by NASA through a grant from the Space Telescope Science Institute, which is operated by the Association of Universities for Research in Astronomy, Inc., under NASA contract NAS 5-26555. S. Benecchi was also supported in part through a Carnegie Fellowship at the Department of Terrestrial Magnetism. We thank Dr. Larry Wasserman and an anonymous reviewer for their helpful comments which improved this manuscript.

## TABLES

TABLE 1. LONG RANGE CANDIDATES

| Name | *a* (AU) | *e* | *i* (deg) | Date of Enc. | Range at Enc. (AU)[c] | Lorri Resolution (km)[b,c] | $\Delta$v required for Enc. (m/s)[c] | Obs arc (days) |
|---|---|---|---|---|---|---|---|---|
| 2011 $HZ_{102}$[a] | 43.2 | 0.008 | 2.4 | 2018/12/16-2018/12/16 | 0.15 - 0.16 | 111-119 | 210-210-210 | 801 |
| 2011 $JY_{31}$[a] | 44.0 | 0.057 | 2.6 | 2018/09/17-2018/09/17 | 0.16 - 0.16 | 118 | 230-230-240 | 801 |
| 2011 $JW_{31}$[a] | 45.6 | 0.094 | 1.9 | 2018/09/27-2018/09/27 | 0.19 - 0.19 | 141 | 280-280-280 | 809 |
| 2011 $HJ_{103}$ | 49.6 | 0.322 | 5.4 | 2017/02/24-2018/01/10 | 0.19 - 0.43 | 141-319 | 530-730-880 | 66 |
| 2013 $LU_{35}$[a] | 43.8 | 0.024 | 3 | 2019/02/04-2019/10/12 | 0.30 - 0.38 | 222-282 | 360-430-450 | 95 |

[a] Designates an object we attempted to observe with HST.

[b] 1 Lorri pixel = 5.959 microradians (Cheng et al. 2008)

[c] The first and last intervals are the 2.5th percentile, 97.5th percentile (the interval enclosing the central 95% of the posterior probability distribution) of 200 samples. The middle value in column 8 is the peak of the probability distribution (the modal velocity).

TABLE 2. OBSERVATIONS

| Object | Image | JD | Calendar Date | Filter | Exptime (s) | Magnitude |
|---|---|---|---|---|---|---|
| 2011 $JW_{31}$ | ibr001faq | 2456188.53241 | 2012/09/18.03241 | F606W | 455 | N/A |
| 2011 $JW_{31}$ | ibr001fbq | 2456188.54088 | 2012/09/18.04088 | F814W | 640 | N/A |
| 2011 $JW_{31}$ | ibr001feq | 2456188.55015 | 2012/09/18.05015 | F814W | 640 | N/A |
| 2011 $JW_{31}$ | ibr001g3q | 2456188.55862 | 2012/09/18.05862 | F606W | 455 | N/A |
| 2011 $JW_{31}$ | ibr002gaq | 2456188.59889 | 2012/09/18.09889 | F606W | 455 | N/A |
| 2011 $JW_{31}$ | ibr002gbq | 2456188.60736 | 2012/09/18.10736 | F814W | 640 | N/A |
| 2011 $JW_{31}$ | ibr002gdq | 2456188.61663 | 2012/09/18.11663 | F814W | 640 | N/A |
| 2011 $JW_{31}$ | ibr002gfq | 2456188.62510 | 2012/09/18.12510 | F606W | 455 | N/A |
| 2011 $JY_{31}$ | ibr011ipq | 2456189.14857 | 2012/09/18.64857 | F606W | 455 | 24.64±0.03 |
| 2011 $JY_{31}$ | ibr011iqq | 2456189.15704 | 2012/09/18.65704 | F814W | 640 | 23.52±0.03 |
| 2011 $JY_{31}$ | ibr011isq | 2456189.20856 | 2012/09/18.70856 | F814W | 640 | 23.63±0.04 |
| 2011 $JY_{31}$ | ibr011iuq | 2456189.21703 | 2012/09/18.71703 | F606W | 455 | 24.69±0.04 |
| 2011 $JY_{31}$ | ibr012iwq | 2456189.28152 | 2012/09/18.78152 | F606W | 455 | 24.66±0.04 |
| 2011 $JY_{31}$ | ibr012ixq | 2456189.28999 | 2012/09/18.78999 | F814W | 640 | 23.83±0.04 |
| 2011 $JY_{31}$ | ibr012izq | 2456189.34353 | 2012/09/18.84353 | F814W | 640 | 23.79±0.04 |
| 2011 $JY_{31}$ | ibr012j1q | 2456189.35200 | 2012/09/18.85200 | F606W | 455 | 24.61±0.04 |
| 2011 $HZ_{102}$ | ibr021dpq | 2456188.21791 | 2012/09/17.71791 | F606W | 455 | 25.68±0.10 |
| 2011 $HZ_{102}$ | ibr021dqq | 2456188.22638 | 2012/09/17.72638 | F814W | 640 | 24.58±0.08 |
| 2011 $HZ_{102}$ | ibr021dsq | 2456188.28014 | 2012/09/17.78014 | F814W | 640 | 24.63±0.09 |
| 2011 $HZ_{102}$ | ibr021duq | 2456188.28861 | 2012/09/17.78861 | F606W | 455 | 25.74±0.09 |
| 2011 $HZ_{102}$ | ibr022dwq | 2456188.35075 | 2012/09/17.85075 | F606W | 455 | 25.72±0.09 |
| 2011 $HZ_{102}$ | ibr022dxq | 2456188.35921 | 2012/09/17.85921 | F814W | 640 | 24.79±0.09 |
| 2011 $HZ_{102}$ | ibr022dzq | 2456188.39891 | 2012/09/17.89891 | F814W | 640 | 24.63±0.09 |
| 2011 $HZ_{102}$ | ibr022e1q | 2456188.40738 | 2012/09/17.90738 | F606W | 455 | 25.66±0.08 |
| 2013 $LU_{35}$ | ibzx01g4 | 2456562.99918 | 2013/09/27.49918 | F606W | 455 | 26.41±0.18 |
| 2013 $LU_{35}$ | ibzx01g5 | 2456563.00765 | 2013/09/27.50765 | F814W | 640 | 25.28±0.15 |
| 2013 $LU_{35}$ | ibzx01g7 | 2456563.01692 | 2013/09/27.51692 | F814W | 640 | 25.19±0.12 |
| 2013 $LU_{35}$ | ibzx01g9 | 2456563.02539 | 2013/09/27.52539 | F606W | 455 | 26.40±0.15 |
| 2013 $LU_{35}$ | ibzx02jg | 2456563.72922 | 2013/09/28.22922 | F606W | 455 | 26.28±0.13 |
| 2013 $LU_{35}$ | ibzx02jh | 2456563.73769 | 2013/09/28.23769 | F814W | 640 | 25.20±0.11 |
| 2013 $LU_{35}$ | ibzx02jj | 2456563.74696 | 2013/09/28.24696 | F814W | 640 | 25.24±0.12 |
| 2013 $LU_{35}$ | ibzx02jl | 2456563.75542 | 2013/09/28.25542 | F606W | 455 | 26.28±0.13 |

TABLE 3. PHOTOMETRY RESULTS

| Object | JD (mid-time) | R (AU) | Δ (AU) | α (°) | $F606W_{avg}$ | $F814W_{avg}$ | F606W-F814W | V-I |
|---|---|---|---|---|---|---|---|---|
| 2011 $JY_{31}$ | 2456189.18280 | — | — | — | 24.67±0.03 | 23.57±0.08 | 1.09±0.08 | 1.37±0.08 |
| 2011 $JY_{31}$ | 2456189.31676 | — | — | — | 24.64±0.04 | 23.81±0.03 | 0.82±0.05 | 1.13±0.05 |
| **2011 $JY_{31,avg}$** | **2456189.24978** | **42.71** | **42.50** | **1.32** | **24.65±0.03** | **23.69±0.16** | **0.96±0.19** | **1.25±0.19** |
| 2011 $HZ_{102}$ | 2456188.25326 | — | — | — | 25.71±0.06 | 24.63±0.06 | 1.11±0.05 | 1.41±0.05 |
| 2011 $HZ_{102}$ | 2456188.37906 | — | — | — | 25.69±0.06 | 24.71±0.11 | 0.98±0.12 | 1.28±0.12 |
| **2011 $HZ_{102,avg}$** | **2456188.31616** | **43.09** | **42.87** | **1.31** | **25.70±0.02** | **24.66±0.07** | **1.04±0.15** | **1.34±0.15** |
| 2013 $LU_{35}$ | 2456563.01229 | — | — | — | 26.41±0.11 | 25.23±0.09 | 1.17±0.07 | 1.50±0.07 |
| 2013 $LU_{35}$ | 2456563.74233 | — | — | — | 26.28±0.09 | 25.22±0.08 | 1.06±0.03 | 1.39±0.03 |
| **2013 $LU_{35,avg}$** | **2456563.37731** | **44.43** | **44.37** | **1.29** | **26.34±0.09** | **25.23±0.02** | **1.11±0.18** | **1.44±0.18** |

## FIGURES

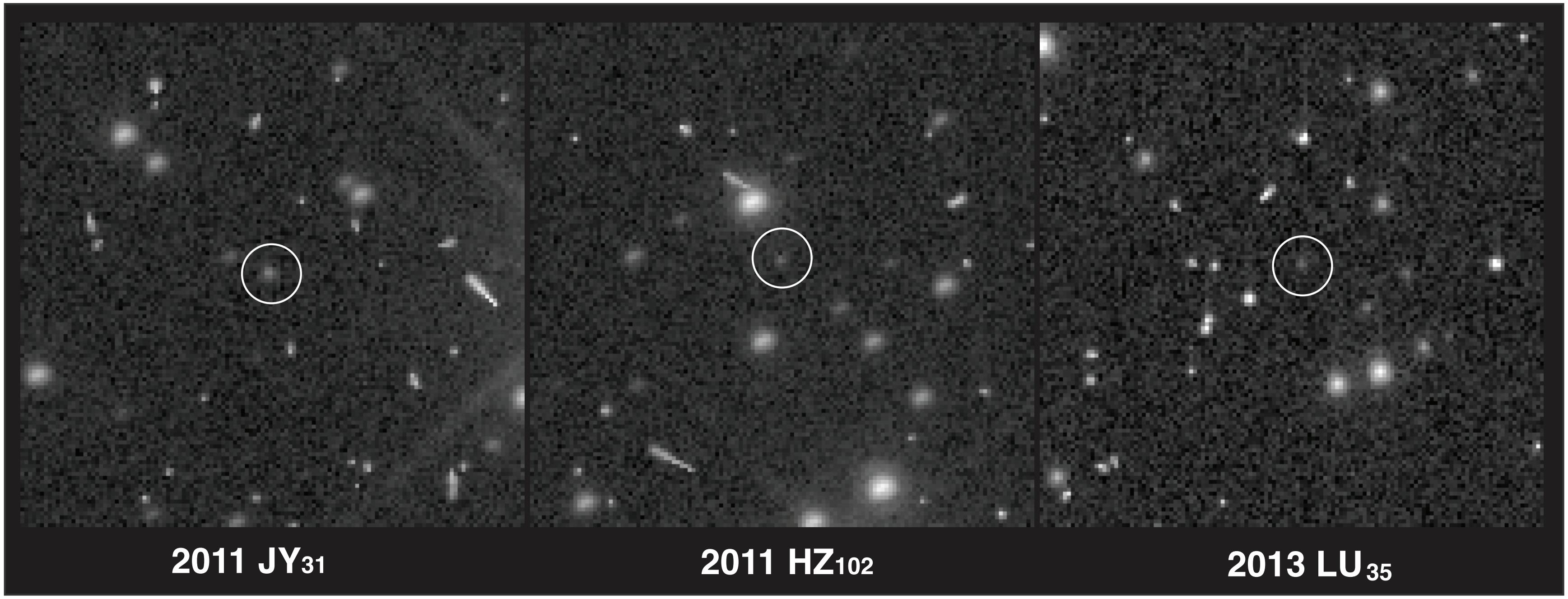

Figure 1. Sample HST images (pipeline calibrated, but otherwise raw format) in the F606W filter for each long-range observation KBO. Magnitudes in this filter ranges from the brightest, 2011 $JY_{31}$ at 24.6 to the faintest, 2013 $LU_{35}$ at 26.3.

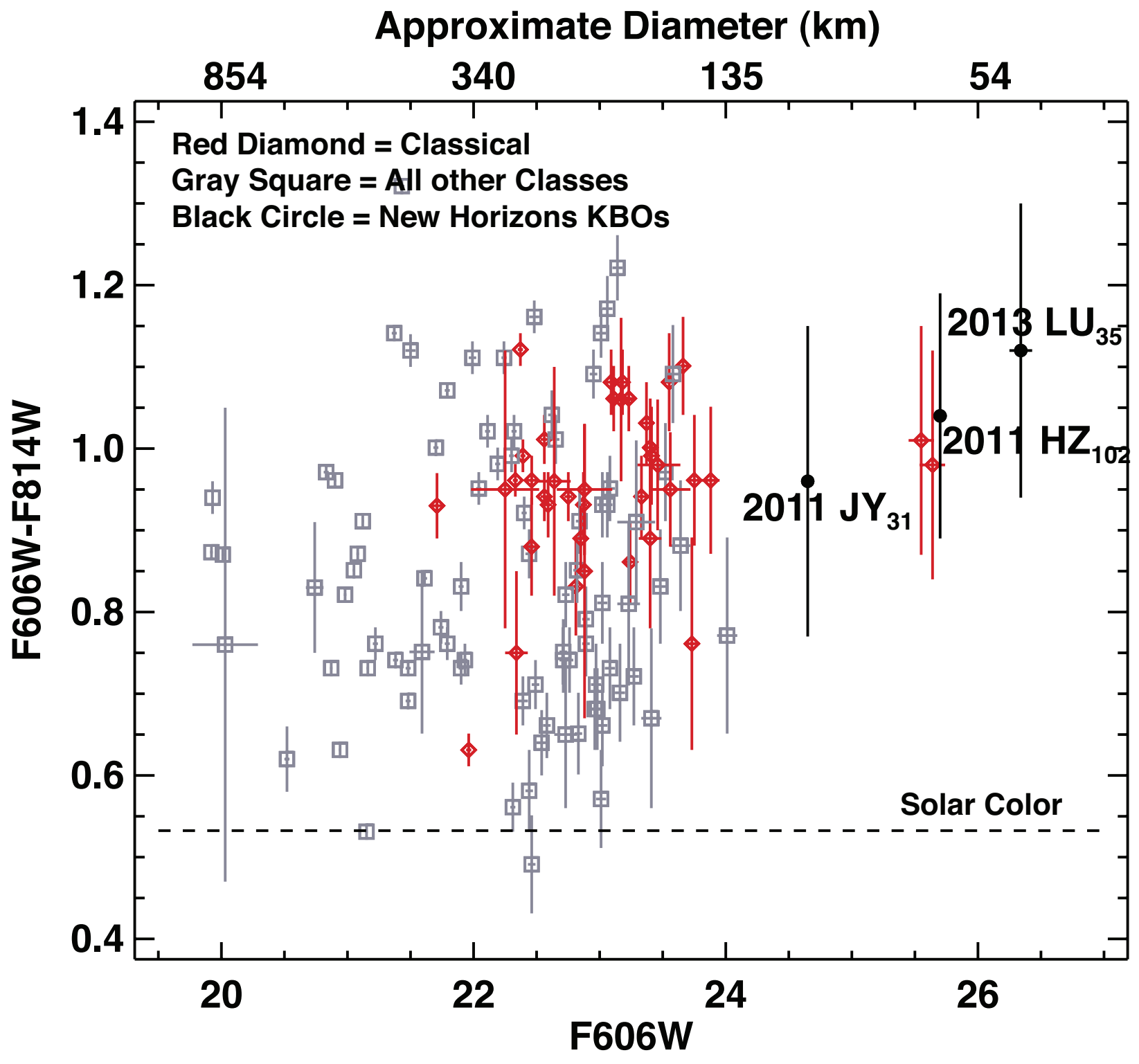

Figure 2. Comparison of New Horizon long-range candidate objects with the color of other KBOs that have been measured in the same HST filters (from Benecchi et al. 2009, Fuentes et al. 2010 and Fraser et al. 2012) with respect to their direct F606W magnitudes. Plotted on the top is an estimate of the effective diameters for these objects following the formulation of Bowell et al. (1989) $d = 10((6.259-0.4 \cdot H_{F606W} - \log \rho)/2)$ km and assuming an albedo, ρ; we assume ρ = 0.1. The F606W-F814W color is similar to ground based Johnson V-Cousins I. Objects plotted as red diamonds are dynamically part of the Cold Classical population, objects in all other classes (Scattered, Resonant, hot Classical and Centaur combined) are plotted as gray squares. The color of the Sun in the HST filter set is plotted as a dashed line. The newly measured objects are plotted as black circles and are fainter by nearly a magnitude or more than the rest of the observed population. These objects are also comparatively in the red portion of the diagram consistent with residing in the cold classical region of the Kuiper Belt.